\documentclass[a4paper]{article}
\makeatletter
\def\@seccntformat#1{\@ifundefined{#1@cntformat}%
   {\csname the#1\endcsname\quad}  % default
   {\csname #1@cntformat\endcsname}% enable individual control
}
\let\oldappendix\appendix %% save current definition of \appendix
\renewcommand\appendix{%
    \oldappendix
    \newcommand{\section@cntformat}{\appendixname~\thesection\quad}
}
\makeatother

\usepackage{color}

\usepackage{amsmath}
\usepackage{amssymb,amsfonts,amsmath}
\usepackage[top=25truemm,bottom=25truemm,left=20truemm,right=20truemm]{geometry}
\usepackage{bm}
\usepackage{cases}
\usepackage[mathlines]{lineno}
\usepackage{comment}
\usepackage{cite}
\usepackage{url}

\usepackage[pdftex]{graphicx} % for arXiv
\usepackage{setspace}
\title{How L\'evy flights triggered by presence of defectors affect \\evolution of cooperation in spatial games}
\bigskip
\author{Genki Ichinose${}^{1*}$, Daiki Miyagawa$^{2}$, Erika Chiba$^{3}$ and Hiroki Sayama$^{4,5}$
\ \\
\ \\
${}^{1}$Department of Mathematical and Systems Engineering, Shizuoka University, \\3-5-1 Johoku, Naka-ku, Hamamatsu, 432-8561, Japan. E-mail: ichinose.genki@shizuoka.ac.jp\\
${}^{2}$Department of Mathematical and Systems Engineering, Shizuoka University, \\3-5-1 Johoku, Naka-ku, Hamamatsu, 432-8561, Japan. E-mail: miyagawa.daiki.18@shizuoka.ac.jp\\
${}^{3}$Graduate School of Informatics, Nagoya University, \\Furo-cho, Chikusa-ku, Nagoya, 464-8601, Japan. E-mail: chiba.erika.j1@s.mail.nagoya-u.ac.jp\\
${}^{4}$Waseda Innovation Lab, Waseda University, Tokyo, Japan \\
${}^{5}$Center for Collective Dynamics of Complex Systems, Binghamton University, \\State University of New York, Binghamton, NY, USA. E-mail: sayama@binghamton.edu\\
$^*$ Corresponding author (ichinose.genki@shizuoka.ac.jp)}

\begin{document}
%\special{papersize=8.5in,11in}
%\setlength{\pdfpageheight}{11in}
%\setlength{\pdfpagewidth}{8.5in}
\maketitle

\section*{Abstract}
Cooperation among individuals has been key to sustaining societies. However, natural selection favors defection over cooperation.
Cooperation can be favored when the mobility of individuals allows cooperators to form a cluster (or group). 
Mobility patterns of animals sometimes follow a L\'evy flight. 
A L\'evy flight is a kind of random walk but it is composed of many small movements with a few big movements.
The role of L\'evy flights for cooperation has been studied by Antonioni and Tomassini. They showed that L\'evy flights promoted cooperation combined with conditional movements triggered by neighboring defectors. However, the optimal condition for neighboring defectors and how the condition changes by the intensity of L\'evy flights are still unclear.
Here, we developed an agent-based model in a square lattice where agents perform L\'evy flights depending on the fraction of neighboring defectors.
We systematically studied the relationships among three factors for cooperation: sensitivity to defectors, the intensity of L\'evy flights, and population density.
Results of evolutionary simulations showed that moderate sensitivity most promoted cooperation.
Then, we found that the shortest movements were best for cooperation when the sensitivity to defectors was high.
In contrast, when the sensitivity was low, longer movements were best for cooperation.
Thus, L\'evy flights, the balance between short and long jumps, promoted cooperation in any sensitivity, which was confirmed by evolutionary simulations.
Finally, as the population density became larger, higher sensitivity was more beneficial for cooperation to evolve. 
Our study highlights that L\'evy flights are an optimal searching strategy not only for foraging but also for constructing cooperative relationships with others.

\section*{Keywords}
L\'evy flight, mobility, evolution of cooperation, spatial game, sensitivity
\clearpage

\section{Introduction}
Cooperative behavior is necessary to sustain human and animal societies \cite{Rand2012TrendsCognSci, Dugatkin1997, Clutton-Brock2009Nature}.
However, the previous studies of evolutionary games show that cooperation is not favored by natural selection compared to defection \cite{Nowak2006Science, Nowak2006}.
Therefore, it has been suggested special mechanisms are needed for cooperation to evolve \cite{Nowak2006Science}.
In the proposed mechanisms, spatial (or network) reciprocity has often been studied \cite{Nowak1992Nature, Santos2005PhysRevLett, Szabo2007PhysRep, Roca2009PhysLifeRev, Ohtsuki2006Nature, Perc2010Biosystems, Perc2013JRSocInterface}.
In those traditional models, individuals do not move in the spatial environment because all spaces are occupied.
Namely, static networks were used for interactions among individuals.
In contrast, many biological organisms have the ability to move. Mobility is a fundamental trait of animals and humans because animals forage for food and people often move when they interact.
Recently, spatial reciprocity with mobility has attracted great attention and various theoretical models have been developed.
Earlier theoretical studies have assumed random (unconditional) movements \cite{Enquist1993AnimBehav, Traulsen2006PNAS, Vainstein2007JTheorBiol, Smaldino2013ChaosSolitonsFractals}.
They revealed that the evolution of cooperation is hindered by mobility because it basically destroys cooperative clusters and leads the population to a well-mixed state.
On the other hand, they also showed cooperation is sustained in the case of low mobility because it contributes to expanding the regions of cooperative clusters.
Recent studies showed that cooperation is enhanced even in the random movements for low mobility if the update rules are properly devised \cite{Sicardi2009JTheorBiol, Antonioni2014JTheorBiol}.

The situation drastically changes if conditional movements have been assumed \cite{Aktipis2004JTheorBiol, Buesser2013PhysRevE, Helbing2009PNAS, Jiang2010PhysRevE, Roca2011PNAS, Tomassini2015JTheorBiol, Ichinose2013SciRep, Tomassini2015JTheorBiol}.
In this case, each agent monitors its current environmental conditions within its local neighborhood and moves to another location if the conditions are found to be undesirable.
In many cases, ``undesirable" refers to the situation that there are many defectors within the local neighborhood.
This contingent movement enhances the evolution of cooperation even if the mobility rate is high because cooperative clusters tend to be created by keeping a distance from defectors.

In those studies, Tomassini and Antonioni focused on a special mobility type, called a L\'evy flight \cite{Tomassini2015JTheorBiol}.
A L\'evy flight is a kind of random walk but it is characterized by many small movements with a few big movements.
More formally, the distance of movements follows a power-law distribution.
It has been shown that some animal species use L\'evy flights when foraging \cite{Viswanathan1999Nature, Viswanathan1996Nature, Sims2012JAnimEcol}.
When resources are randomly distributed and there is no information on their locations, a search pattern based on a  L\'evy flight type is optimal \cite{Lomholt2006PNAS}.
Another study shows that humans also use L\'evy flights \cite{Brockmann2006Nature}.

Tomassini and Antonioni studied the evolution of cooperation in spatial games where agents perform L\'evy flights \cite{Tomassini2015JTheorBiol}.
In the model, they assumed two types of conditions where L\'evy flights are performed by agents: 1) Agents always perform L\'evy flights, 2) agents perform L\'evy flights only when more than half of their neighbors are defectors.
They showed that cooperation evolved only in the latter case. 

Motivated by this study, we focus on the evolution of cooperation of mobile agents that perform L\'evy flights in spatial games.
Tomassini and Antonioni's model was a bit extreme in the sense that they only consider two types of conditions for L\'evy flights.
Here, we consider a continuous range of sensitivity to the presence of defectors to identify the optimal level of L\'evy flights for the evolution of cooperation in spatial games.
From another perspective, it was unknown whether the power-low characteristic of movement promoted cooperation.
Simply, big jumps which are not due to L\'evy flights may lead to the evolution of cooperation or short jumps are better than such big jumps.
To test this case, here we also let the intensity of L\'evy flights be adjustable in the model.
As the special case, uniform movements where agents can move to any space regardless of distances with equal probability and constant movements where agents always move a fixed distance are also realized.
Through this extension, we study how the intensity of L\'evy flights affects cooperation. Finally, we reveal how the sensitivity which yields the optimal cooperation changes depending on the population density.

\section{Model}
We previously developed an agent-based model of the evolution of cooperation in a square lattice where the sensitivity to neighboring defectors in L\'evy flights is adjusted by step functions \cite{Ichinose2020ALIFE}.
Here, we extended it so that the intensity of L\'evy flights can be adjusted. 
First, agents are randomly distributed into an $L\times L$ lattice.
The density of the agents is given by $\rho$. Thus, the number of agents is $N=L^2 \rho$.
At the beginning of a simulation, half of the agents are cooperators and the other half are defectors.
Then, the following process is repeated until the specified number of time steps ($t_{\rm end}=500$) is obtained.

\begin{figure}[tb]
	\centering
	\includegraphics[width=100mm]{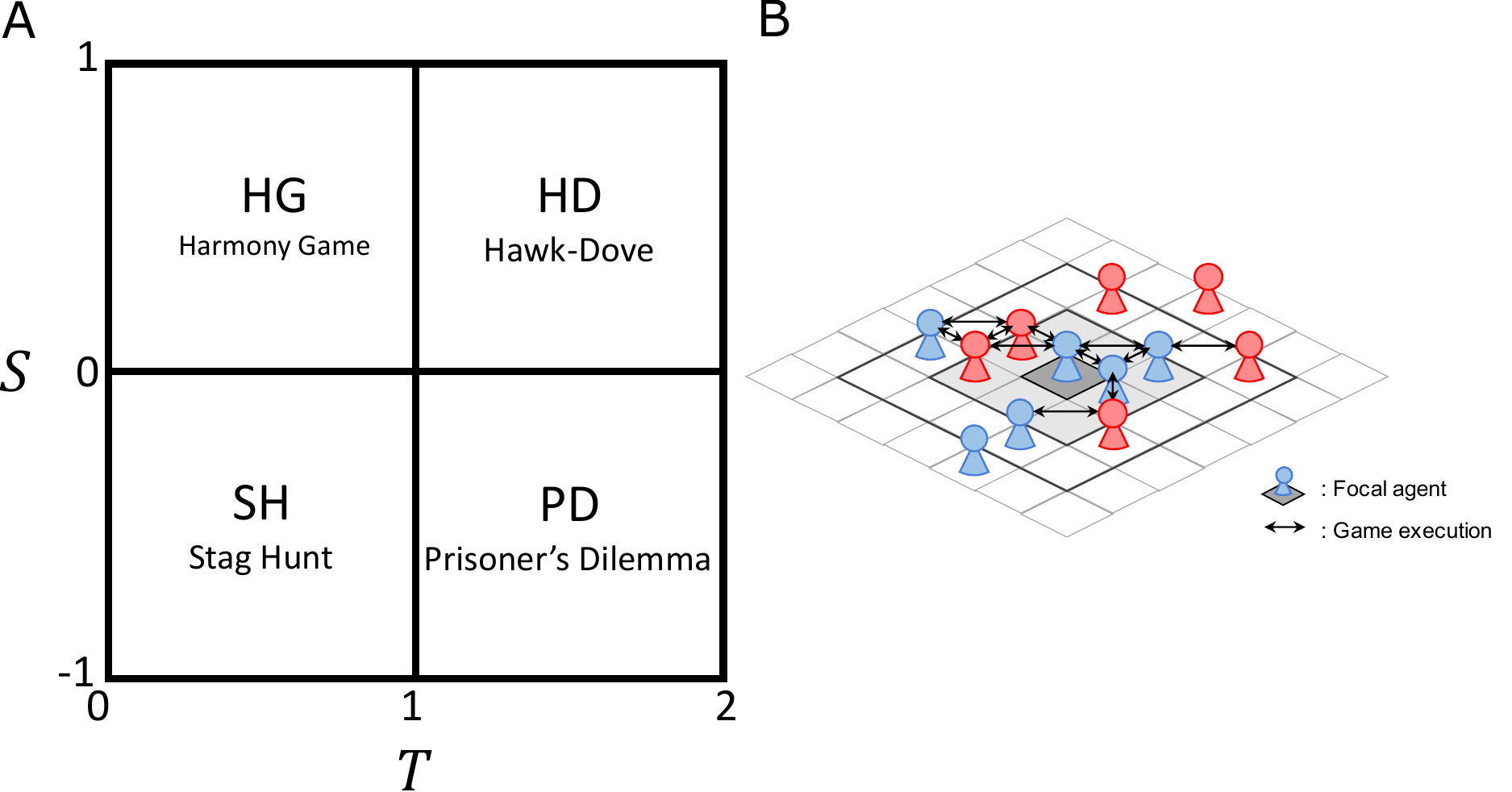}
	\caption{(A) Four games used in the model. (B) Game executions in Step 2. The figure shows an example.
	The focal agent (center) plays the game with its neighbors (two cooperators and two defectors) and obtains the payoff.
	The neighbors also play the game with their neighbors and obtain payoffs.}
	\label{fig:Step2}
\end{figure}

\renewcommand{\labelenumi}{\arabic{enumi}. }
\begin{enumerate}
	\setlength{\itemsep}{0pt}
	\setlength{\itemindent}{5pt}
	\setlength{\labelsep}{0pt}
\item One agent is randomly selected from the whole population. (This agent may be selected multiple times in one time step because we used an asynchronous update scheme.)
\item The agent (located in the center in Fig.~\ref{fig:Step2}B) plays one of four games (Fig.~\ref{fig:Step2}A) with its neighbors and obtains the payoff. The neighboring agents (located in the light gray area in Fig.~\ref{fig:Step2}B) also play the game with their neighbors and obtain payoffs. The detail of these games is described below.
\item The agent imitates the strategy of the neighbor that obtained the highest payoff within the neighborhood, including itself. If two or more agents have the highest payoff at the same time, the agent randomly picks up and imitates the strategy of one of those agents.
\item The agent is unsatisfied when the neighbors are defectors. If the fraction of defectors is equal to or greater than a threshold value, it performs a L\'evy flight to another cell if the cell is empty. Otherwise, the agent does not move.
\item The above is repeated $N$ times, which is regarded as one time step ($t$).
\end{enumerate}

In Step 2, the agent and its opponent play one of four common two-person, two-strategy, symmetric games (Fig.~\ref{fig:Step2}A).
The two strategies are cooperation (C) and defection (D).
If both cooperate, they receive $R$; if one cooperates and the other defects, the former obtains $S$ and the latter obtains $T$; if both defect, they receive $P$. 
The games are classified into the following four depending on the payoff relationships: the Harmony Game (HG); $R>T>P$ and $R>S>P$ \cite{Licht1999YaleJIntLaw}, the Stag Hunt (SH); $R>T>P>S$, the Prisoner's Dilemma (PD); $T>R>P>S$, and the Hawk-Dove game (HD); $T>R>S>P$.
In the PD, cooperation is the strategy that benefits others by paying costs while defection is the strategy that enjoys the benefits from cooperators without paying any cost.

In Step 4, the condition of dissatisfaction is provided as follows.
We define sensitivity $s = 1 - \frac{i}{n_{{\rm max}}} \, (0 \leq s \leq 1)$ where $n_{\rm max}$ denotes the maximum number of agents in the neighborhood, that is, $n_{\rm max}=8$.
$i$ denotes a threshold value for every level of sensitivity.
We consider nine threshold values $i=0, 1, ..., 8$.
Then, we assume the following step functions which decide whether agents perform L\'evy flights or not
\begin{equation}
P(s) =  \begin{cases}
    1 & (1-\frac{n_{\rm D}}{n} \leq s) \\
    0  & (\mathrm{otherwise}),
  \end{cases}
  \label{eq:levy_step}
\end{equation}
where $n$ denotes the number of agents in the neighborhood and $n_{\rm D}$ denotes the number of defectors in the neighborhood.
$P(s)$ is the probability that agents perform L\'evy flights.
From Eq.~\ref{eq:levy_step}, nine step functions are obtained.
When the first equation in Eq.~\ref{eq:levy_step} is satisfied, agents perform L\'evy flights. Note that all agents have the same sensitivity.

The jump distance of L\'evy flights, $x$, is given by a power-law distribution $P(x)=Cx^{-\alpha}$, where $C$ is a normalization constant such that $C \sum_{x=1}^{L} x^{-\alpha}=1$.
Note that as we use periodic boundary conditions, if the jump of an agent is over a boundary, the agent comes back from the
opposite boundary. Thus, the maximum jump length is $\lfloor L/2 \rfloor$ rather than $L$. If jump length $x$ is larger than $\lfloor L/2 \rfloor$, it is equal to $L - x$.
The probability distribution $P(x)$ is affected by this spatial periodicity, but only very slightly, because the probabilities of such long-range jumps are very small.
We systematically varied $\alpha$ in the range of $0 \leq \alpha \leq 10$ in the simulations.
$\alpha=0$ can be considered as the special case. In this case, an agent moves to another cell regardless of the distance with equal probability.
In other words, an agent moves to another cell based on a uniform distribution.
Thus, we call this special case a ``uniform movement" hereafter.
Actual observations suggest that animals use $1 \leq \alpha \leq 3$.
As the control experiments, we also considered the cases of contingent jumps to certain distances.
The settings are $P(x)=1, x=1, 2, 3, 5, 10$ where agents always jump a certain distance (1, 2, 3, 5 or 10).

We use $L=50$ and $\rho=2/3$ unless otherwise noted.
For the game parameters, we fix $(R, P)=(1,0)$ while changing $-1 \leq S \leq 1$ and $0 \leq T \leq 2$. 

\section{Result}
\subsection{L\'evy flights promote cooperation}
\begin{figure*}[tb]
	\centering
	\includegraphics[width=\textwidth]{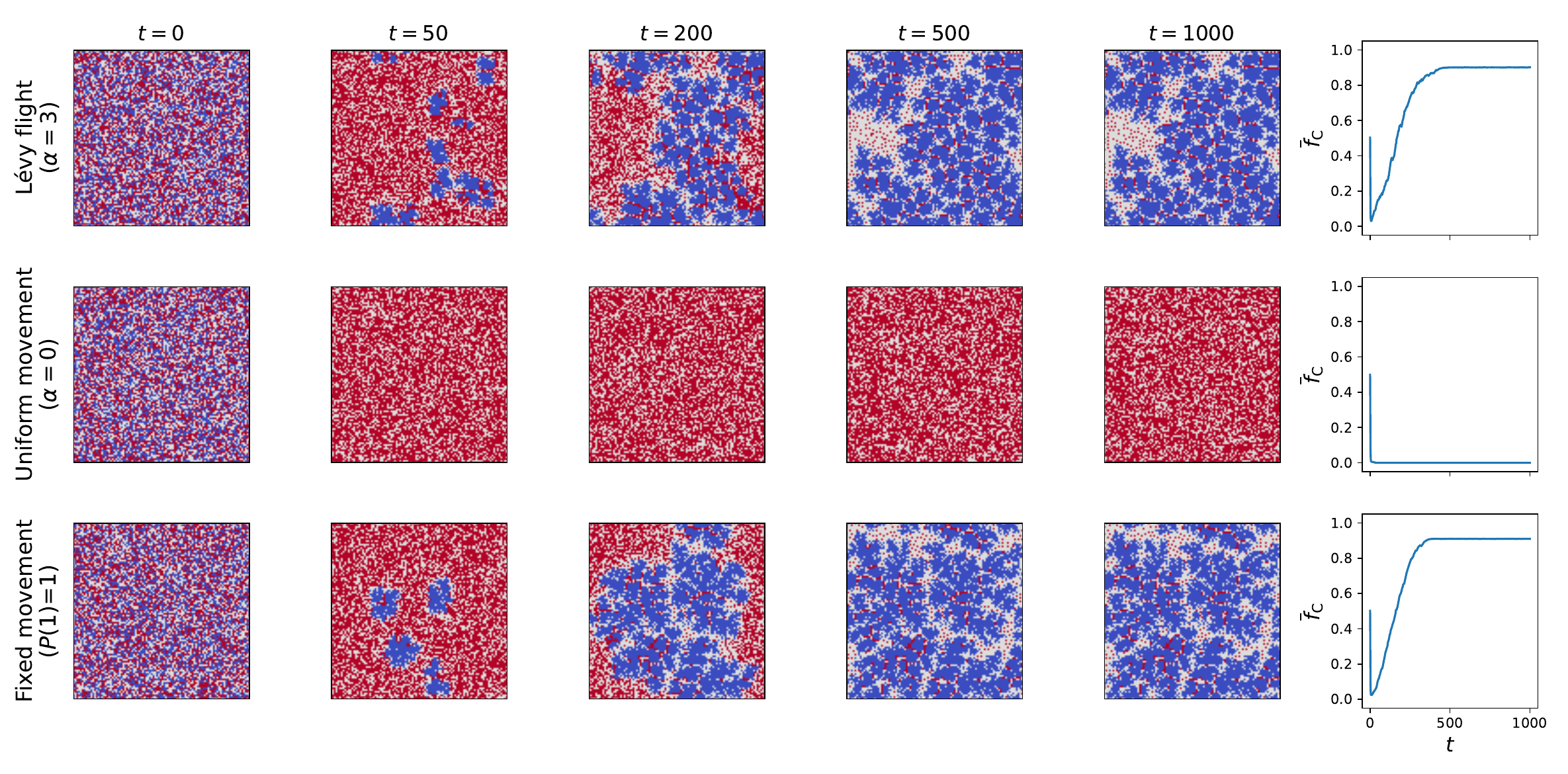}
	\caption{Snapshots of a simulation. Cooperators (Defectors) are shown in blue (red). Initially ($t$ = 0), cooperators and defectors are fifty-fifty. When uniform movements ($\alpha=0$) are assumed, cooperators die out. In contrast, cooperators spread when L\'evy flights or fixed movements are assumed. At $t=50$, cooperators almost go extinct but a few cooperative clusters survive in the two cases. Finally, the cluster of cooperators can invade the sea of defectors. We used the PD game where $(R,S,T,P)=(1,-0.4,1.4,0)$. $L=100$ and $\rho=2/3$.}
	\label{fig:snapshots}
\end{figure*}

First, we focus on whether and how L\'evy flights promote cooperation in spatial games.
Figure~\ref{fig:snapshots} shows the snapshots of the simulation where the sensitivity is $s=1/2$ and three distinct cases $\alpha=3.0$ (L\'evy flight), $\alpha=0$ (uniform movement), and $P(1)=1$ (fixed movement) are compared.
Here, we set $(S, T)=(-0.4, 1.4)$, thus the game is the PD.
In the figure, cooperators (defectors) are shown in blue (red).
We also provide the whole simulation as a video\footnote{\url{https://doi.org/10.6084/m9.figshare.14827134.v2}}.

Cooperators die out in this parameter setting when uniform movements ($\alpha=0$) are assumed.
This is because cooperative clusters are not maintained as agents are well-mixed by uniform movements ($\alpha=0$). 
In contrast, cooperators eventually spread in the form of clusters when L\'evy flights ($\alpha=3.0$) or fixed movements ($P(1)=1$) are assumed.
Even in these two cases, in the early stages of the simulation, cooperators almost go extinct but a few clusters still survive (from $t=0$ to $50$).
If cooperators are clustered, they can obtain higher payoffs within the areas.
Thus cooperative clusters can survive.
Then, agents at the borderlines tend to imitate the cooperative strategy because the payoff of an agent in a cooperative cluster is high.
Therefore, cooperative clusters gradually expand their regions.

We now focus on the difference between L\'evy flights ($\alpha=3.0$) and fixed movements ($P(1)=1$).
The dynamics of strategies change in the fixed movements ($P(1)=1$) is a bit faster than that in the L\'evy flights ($\alpha=3.0$) (See also the video to confirm it).
In the fixed movements ($P(1)=1$), agents only move to cells with a distance of one.
In this case, the dynamics are faster because the strategy change settles down quickly from cooperation to defection or vice versa.
In contrast, in the case of the L\'evy flights ($\alpha=3.0$), cooperators and defectors are sometimes mixed again by rare big jumps, which contribute to a longer time for the conversion.
However, there is no difference between L\'evy flights ($\alpha=3.0$) and fixed movements ($P(1)=1$) in the final fraction of cooperators.
In L\'evy flights ($\alpha=3,0$), the probability that agents to a cell at a distance of one are about 0.8320.
Thus, we find that the shortest movements prominently contribute to maintaining cooperative clusters and these movements are useful to gradually expand those clusters.
%Then, cooperators can expand their areas by moving locally and avoiding defectors based on the adaptive L\'evy flights (from $t=200$ to $500$).
%Moreover, L\'evy flights which consist of rare big movements also benefit cooperation because cooperators can inhabit new areas.

\begin{figure*}[hbtp]
	\centering
	\includegraphics[width=\textwidth]{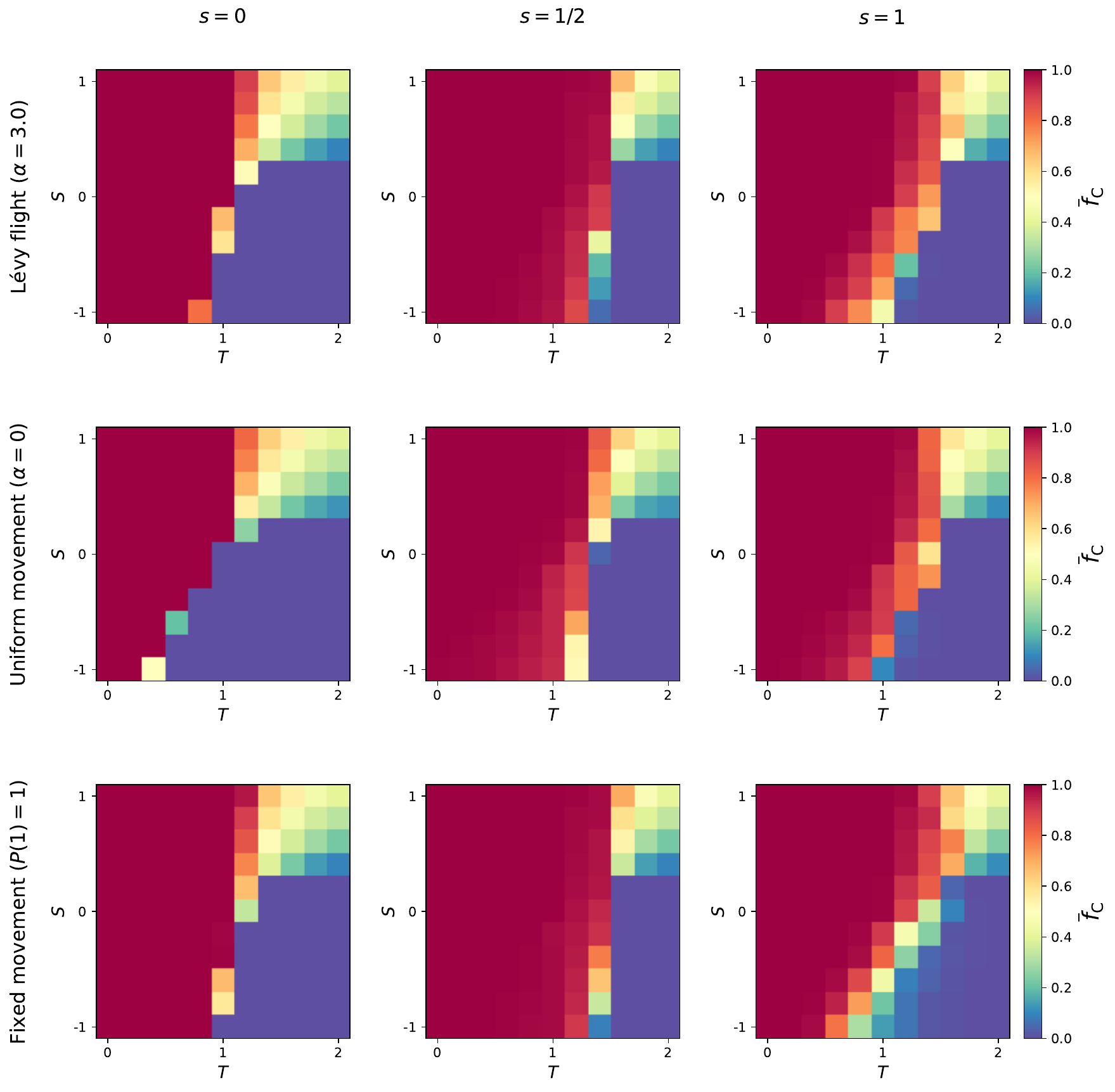}
	\caption{Fraction of cooperators $\bar{f}_{\rm C}$ in the whole $TS$ plane when $s=0, 1/2,$ and $1$. The plane is divided into the four games (HG, SH, PD, and HD) depending on the $T$ and $S$ values. We averaged 10 simulation runs for each data point.}
	\label{fig:fc}
\end{figure*}

Next, we show how cooperation evolved in the whole $TS$ plane when $s=0, 1/2,$ and $1$.
Figure \ref{fig:fc} shows the average fraction of cooperators, denoted by $\bar{f}_{\rm C}$, at the final step of the simulations ($t_{\rm end}=500$).
Here, $\alpha=3.0$, $\alpha=0$, and $P(1)=1$ are compared again.

For all cases, cooperation evolved when the games were the HG and the SH because cooperation between two agents ($R$) is most beneficial.
In contrast, cooperation was hard to evolve when the games were the PD and the HD.
In those two games, unilateral defection ($T$) is most beneficial.
Moreover, defection is the dominant strategy in the PD due to $T>R$ and $P>S$.
Thus, the PD resulted in the worst case for cooperation to evolve.
When we compare the three results for $s$ values, cooperation evolved in the moderate sensitivity $s=1/2$.
%Finally, we focus on the difference \add{among $\alpha=3.0$, $\alpha=0$, and $P(1)=1$}. Overall, cooperation is promoted more in the \add{cases of $\alpha=3.0$ and $P(1)=1$} among the three sensitivities.
By this time, we do not see any characteristic difference in the results between  $\alpha=3.0$ and $P(1)=1$.
We discuss these results in detail and clarify the difference between  $\alpha=3.0$ and $P(1)=1$ in the next section by showing all $\alpha$ and $s$ values.

\subsection{Optimal sensitivity for cooperation}
\begin{figure}[hbtp]
	\centering
	\includegraphics[width=\columnwidth]{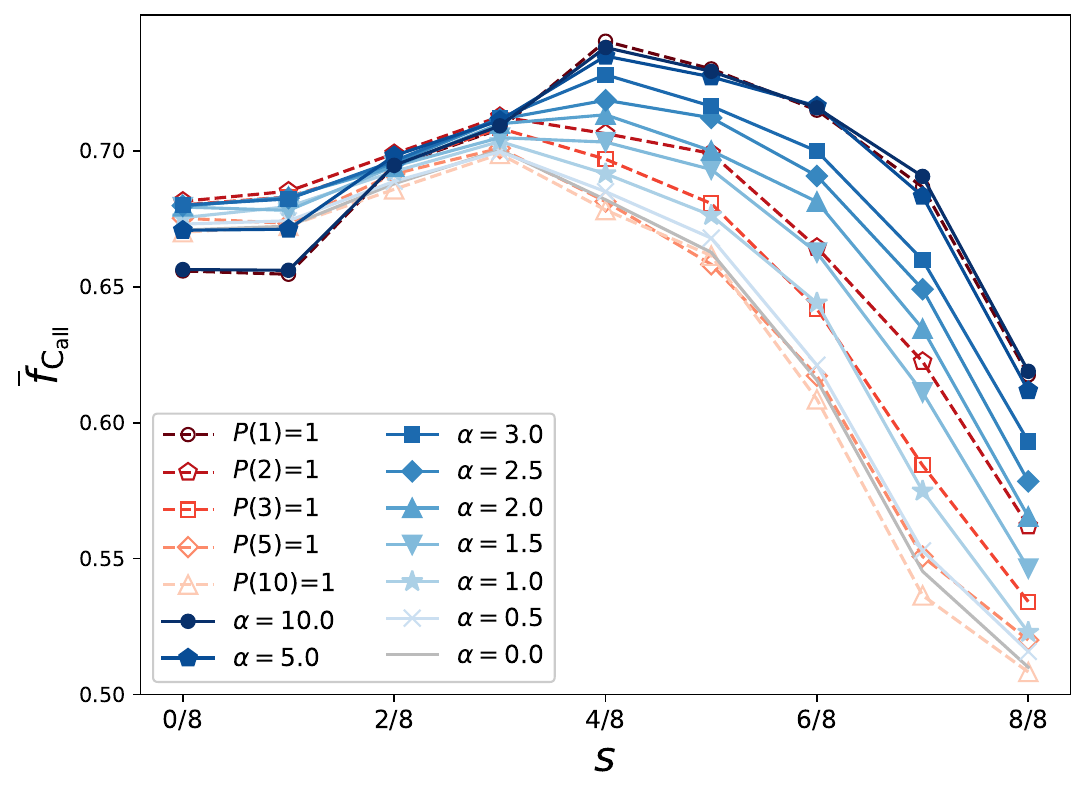}
	\caption{Fraction of cooperators $\bar{f}_{\rm C_{\rm all}}$ as a function of sensitivity $s$ when $\alpha$ is varied where $\bar{f}_{\rm C_{\rm all}}$ is obtained by averaging all $\bar{f}_{\rm C}$ in the whole parameter ranges ($0 \leq T \leq 2$ and $-1 \leq S \leq 1$). For each point on the lines, 10 simulation runs are averaged. Cooperation was promoted the most in the moderate sensitivity $s=1/2$.}
	\label{fig:sensitivity-fC}
\end{figure}

Here, we try to find out which $s$ produces the optimal cooperation level when the intensity of L\'evy flights $\alpha$ is varied.
We changed $s$ with summing up $-1 \leq S \leq 1$ and $0 \leq T \leq 2$.
Figure \ref{fig:sensitivity-fC} shows the optimal sensitivities depending on $\alpha$.

We first found that the moderate sensitivity most promoted cooperation in all cases (Fig.~\ref{fig:sensitivity-fC}).
We explain the results by dividing them into two cases depending on $s$ values. When $s \ge 1/2$, $P(1)=1, \alpha=5.0$, and $\alpha=10.0$ were best for cooperation.
Because the probability that agents jump to a distance of one is 0.9644 for $\alpha=5.0$, 0.9990 for $\alpha=10.0$, and 1 for $P(1)=1$, it implies that the shortest movements promote cooperation when $s \ge 1/2$.
When $s$ is high, agents frequently move in the presence of neighboring defectors.
In such cases, because big jumps destroy cooperative clusters, short movements are needed to maintain these clusters.
On the other hand, when $s \le 1/4$, $P(2)=1, \alpha=2.5$, and $\alpha=3.0$ were best for cooperation.
These low $s$ values mean that agents are patient with neighboring defectors.
In this case, cooperative clusters are invaded by defectors until the forms of clusters collapse.
Thus, even if cooperators jump to close cells, it is difficult for them to reform clusters because there are fewer cooperators nearby.
In those situations, there is a possibility that cooperative clusters are reorganized in different places due to a bit further jumps ($P(2)=1$) or rare big jumps ($\alpha=2.5, 3.0$), which work better for cooperation.

The total results mean that L\'evy flights are not particularly better than fixed movements.
Is there no advantage for rare big movements?
From the results, it seems that there is merit for L\'evy flights.
That is, L\'evy flights with $\alpha=2.5$ and 3.0 sufficiently promote cooperation regardless of $s$.
To investigate this conjecture, in the next section, we conducted evolutionary simulations where the intensity of L\'evy flights is an evolvable trait as well as the strategy of agents.

\subsection{Evolution of the intensity of L\'evy flights}
In the previous subsection, we realized that L\'evy flights were not so optimal when $s$ was fixed.
Instead, L\'evy flights show a better performance in any $s$ value.
Therefore, we study what types of mobility promote cooperation when $s$ is diverse.
Here, we consider mobility as one of the evolvable traits.
In addition to a strategy in the game, each agent has $\alpha_i$ and $\beta_i$ related to the probability of jump distance $x$ in the following equation,

\begin{equation}
P(x, \alpha_i, \beta_i) \sim (|x - \beta_i|+1)^{-\alpha_i}.
\end{equation}
Note that when $\beta_i = \beta = 1$ and $\alpha_i = \alpha = \rm{const.}$ (all agents have the same $\alpha$ and $\beta$), the situation corresponds to the original model we used so far.
Also, if $\beta$ converges to $x'$ while $\alpha \to \infty$ by evolution, it means that all agents move to a fixed distance $x'$ the same as $P(x')=1$.
We set $(R, S, T, P)=(1, -0.3, 1.2, 0)$ as it represents the PD game.
At the beginning of a simulation, half of the agents are cooperators and the other half are defectors and we randomly set integers with the ranges of $\alpha_i \in [0, 10]$ and $\beta_i \in [1, 11]$ for each agent.
As for $s$, randomly selected $s_i\; (0\leq s_i \leq 1)$ is assigned to each agent at every time step.

\begin{figure}[hbtp]
	\centering
	\includegraphics[width=\columnwidth]{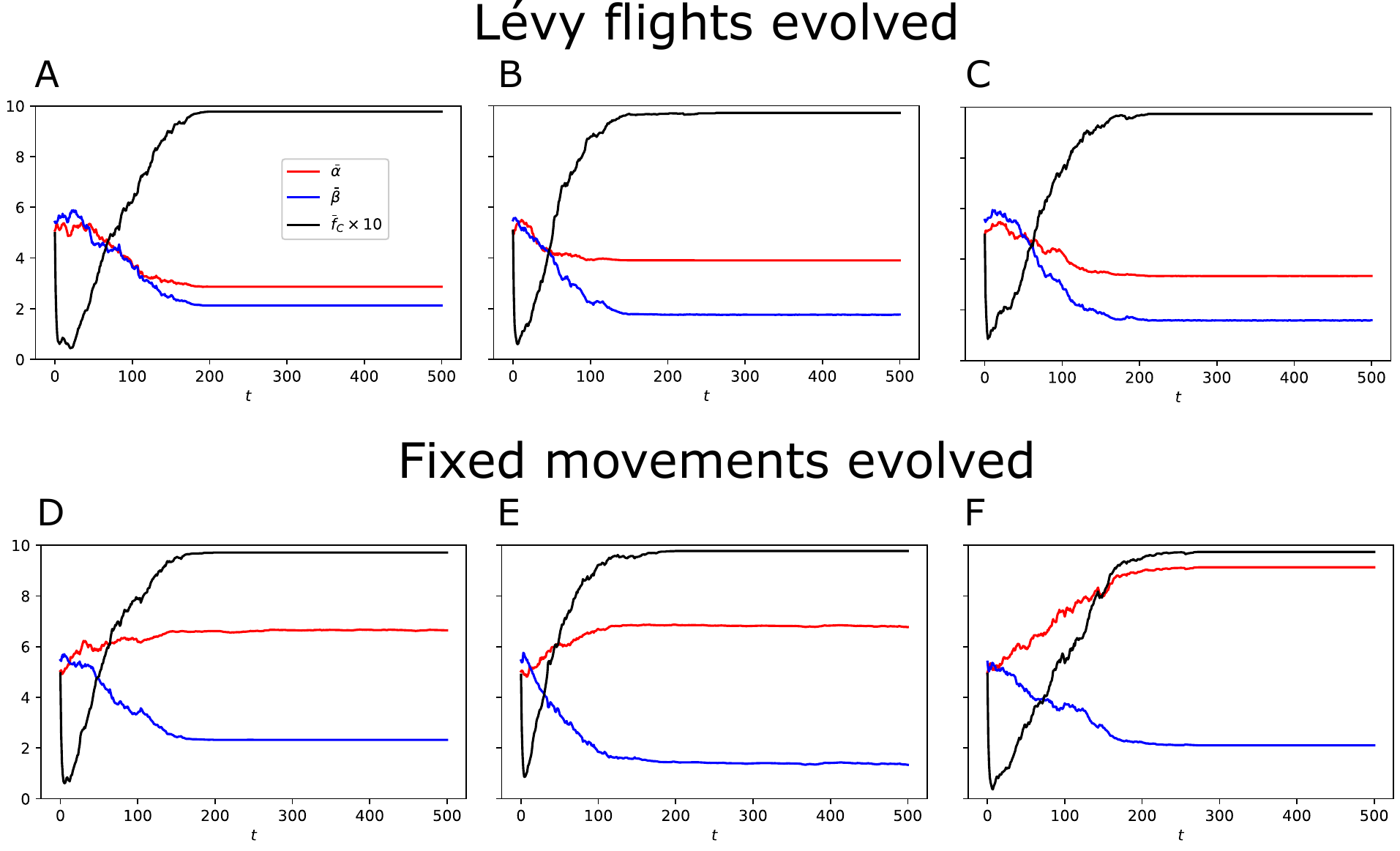}
	\caption{Average of $\alpha_i$ ($\bar{\alpha}$), $\beta_i$ ($\bar{\beta}$), and the fraction of cooperators ($\bar{f}_{\rm C}$) over time.  $\bar{f}_{\rm C}$ is multiplied by ten to align with the other two lines. Six results are shown as examples. PD game with $(R, S, T, P)=(1, -0.2, 1.2, 0)$ was used. $s_i$ is randomly assigned to each agent every time step.}
	\label{fig:evo}
\end{figure}

We conducted thirty evolutionary simulations in total. 
All of the results are provided as supplementary information (Fig.~S1).
Here, we pick up six results as examples shown in Fig.~\ref{fig:evo}.
Figure \ref{fig:evo} shows the average of $\alpha_i$ ($\bar{\alpha}$), $\beta_i$ ($\bar{\beta}$), and the fraction of cooperators ($\bar{f}_{\rm C}$) over time by evolution.
The results are classified into three cases.
The first case was that L\'evy flights evolved.
We say that L\'evy flights evolved when $\bar{\alpha}$ converged lower than 5 and $\bar{\beta}$ converged lower than 5.5.
This case happened $9/30 = 30\%$.
Panels (A), (B), and (C) in Fig.~\ref{fig:evo} mean that L\'evy flights are achieved by evolution because $\alpha \approx 2.87$ and $\beta \approx  2.13$ for (A), $\alpha \approx 3.90$ and $\beta \approx 1.77$ for (B), and $\alpha \approx 3.34$ and $\beta \approx 1.59$ for (C) are obtained.
The second case was that fixed movements evolved.
We say that fixed movements evolved when $\bar{\alpha}$ converged larger than 5 and $\bar{\beta}$ converged lower than 5.5.
This case happened $17/30 \approx 57\%$.
Panels (D), (E), and (F) show such cases.
In panel (D), $\alpha \approx 6.64$ and $\beta \approx 2.32$ are obtained. In panel (E), $\alpha \approx 6.77$ and $\beta \approx 1.34$ are obtained.
In panel (F), $\alpha \approx 9.13$ and $\beta \approx 2.11$ are obtained.
These results are close to the case of fixed movements $P(1)=1$ or $P(2)=1$ because agents almost always jump to a cell at a distance of one or two in the two results due to large $\bar{\alpha}$.
The last case was that evolution favored defection over cooperation.
This case happened $2/30 \approx 7\%$.
In this case, all agents became defectors.
Thus, because there is no selection pressure on $\alpha$ and $\beta$, these values fluctuate by random drift.
See Fig.~S1.
In the other two simulations, evolution did not lead to convergence within 500 time steps.

In this way when $s$ is diverse, L\'evy flights and cooperation often co-evolve by evolution.
It is known that L\'evy flights are the optimal balance between exploitation and exploration \cite{Murakami2019JRSocInterface} and they have functional advantages near a critical point \cite{Abe2020PNAS}.
In our model, the optimal balance was effective to the sensitivity to defectors, which was why L\'evy flights evolved.

\subsection{Change of optimal sensitivities depending on densities}
\begin{figure}[hbtp]
	\centering
	\includegraphics[width=\columnwidth]{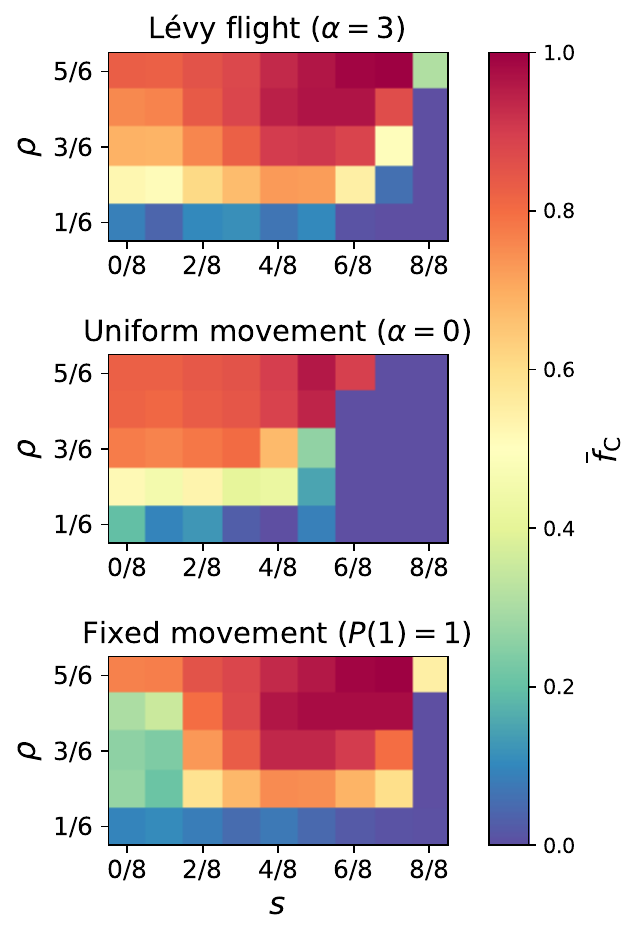}
	\caption{Fraction of cooperators $\bar{f}_{\rm C}$ as functions of densities and sensitivity. Top: L\'evy flights ($\alpha=3.0$). Center: Uniform movements ($\alpha=0.0$). Bottom: Fixed movements ($P(1)=1$).
			PD game with $(R, S, T, P)=(1, -0.4, 1.2, 0)$ was used. 
				We averaged 10 simulation runs for each data point. }
	\label{fig:density}
\end{figure}

Finally, we focus on how cooperation evolves depending on density $\rho$. L\'evy flights ($\alpha=3.0$), uniform movements ($\alpha=0.0$), and fixed movements ($P(1)=1$) are compared.
Figure \ref{fig:density} shows $\bar{f}_{\rm C}$ when sensitivity $s$ and density $\rho$ were changed.
As seen in the figure, L\'evy flights (with $\alpha=3.0$) and fixed movements ($P(1)=1$) promoted cooperation compared to uniform movements as a whole (red regions in the first and third panels in Fig.~\ref{fig:density} are larger than that in the second panel). As explained above, this is because the shortest movements have the advantage of forming cooperative clusters.

Next, we examine the effect of density on cooperation.
When the sensitivity was at its highest $s=1$, cooperation did not evolve at all in all cases.
When the sensitivity was too low $s \leq 1/4$, cooperation did not evolve much.
Thus, even when the density was changed, moderate sensitivities $1/4 \leq s < 7/8$ were best for cooperation to evolve.

Moreover, as the density became larger, higher sensitivity promoted more cooperation in all cases.
In sparse situations (low densities), cooperative clusters tend to be maintained because they are surrounded by few defectors.
In contrast, in dense situations, cooperative clusters tend to be destroyed by surrounding defectors.
In that case, it is better for cooperators to escape from their current positions by moving to other cells.
Thus, higher sensitivity can promote cooperation in dense situations.

\section{Conclusion}
We investigated the effect of how sensitivity to defectors when performing L\'evy flights promotes the evolution of cooperation.
We constructed an agent-based model where agents play games with their neighbors, update their strategies, and perform L\'evy flights to move to other cells in a square lattice. 
Compared to the previous work, we tested various levels of sensitivity to defectors for the condition of L\'evy flights and analyzed the relationship between the sensitivity and density for cooperation.
We also checked how the intensity of L\'evy flights affected cooperation by changing $\alpha$.
The evolutionary simulations showed the following facts.
First, cooperation was most promoted in the moderate sensitivity.
Second, the optimal movements were different depending on the sensitivity.
When the sensitivity was high, the shortest movements where agents jump to a cell with a distance of one were best for cooperation.
On the other hand, when the sensitivity was low, a bit further or rare big jumps were best for cooperation.
Our results implied that L\'evy flights were not so optimal for promoting cooperation when the sensitivity was fixed.
However, through evolutionary simulations of the intensity of L\'evy flights, we found that agents evolved to use L\'evy flights when the sensitivity was diverse.
Finally, as the density increased, higher sensitivity to defectors was better for cooperation to evolve. 

We previously suggested that big jumps promoted cooperation in spatial games \cite{Ichinose2013SciRep}.
However, in that study, agents tended to move to distant cells as the fraction of neighboring defectors became high.
Namely, the distances of jumps were the functions of neighboring defectors.
Therefore, a relatively high cognitive ability was required because agents had to have not only the detection of neighboring environments but also the desire which kept them at a distance from defectors.
Contrary to this previous work, in this study, mobile agents only require a simple cognitive ability which is the detection of neighboring environments because jump distances do not depend on the fraction of the defectors.
Even in this case, we showed that cooperation sufficiently evolved.
Thus, our study highlights the possibility of the evolution of cooperation in biological mobile organisms which have simple cognitive abilities.
We can include a desire for agents to keep a distance from defectors into the current model, which is one direction of future work.

L\'evy flights are known as an optimal search strategy when the targets (e.g., food, mates, or habitats) are sparsely distributed in the environment.
Here, we showed that L\'evy flights are effective not only for those targets but also the sensitivity to defectors in the context of the evolution of cooperation.
This is a new finding that we first discovered.

\section{Acknowledgments}
This work is supported by HAYAO NAKAYAMA Foundation for Science \& Technology and Culture.
%This work was supported by NSF grant No.\ PHY-9723972.

\end{document}